\documentstyle[floats,aps]{revtex}

\input epsf         
\epsfverbosetrue    
\def\postscript#1{\begin{center}\leavevmode
\hbox{\epsfxsize=0.95\columnwidth\epsfbox{#1}}\end{center}}

\begin{document}

\twocolumn[\hsize\textwidth\columnwidth\hsize\csname@twocolumnfalse%
\endcsname

\draft

\title{Absence of anisotropic universal transport in YBCO}

\author{W. C. Wu}
\address{Department of Physics, National Taiwan Normal University,
Taipei 11718, Taiwan}

\author{J. P. Carbotte}
\address{Department of Physics and Astronomy, McMaster University\\
Hamilton, Ontario, Canada L8S 4M1}

\date{\today}

\maketitle

\begin{abstract}
There exists significant in-plane
anisotropy between $a$ and $b$ axis for various properties in YBCO.
However recent thermal conductivity measurement by Chiao {\em et al.}
which confirms previous 
microwave conductivity measurement by Zhang {\em et al.},
shows no obvious anisotropy in the context of universal transport. 
We give a possible explanation of why the anisotropy is seen in 
most properties but not seen in the universal transport.
\end{abstract}

]

\narrowtext

The topic of anisotropic superconductors is currently
of great interest in relation to high-temperature superconductivity.
For example, in the most studied material YBCO,
due to the orthorhombic nature of its band structure,
it exhibits strong in-plane anisotropy between some measurable
quantities along the $a$ and $b$ axis.
Anisotropic behavior is confirmed in the 
magnetic penetration depth $\lambda$ \cite{Zhang,Basov},
normal-state resistivity $\rho$\cite{Friedmann90}, 
optical conductivity $\sigma_s(\omega)$ \cite{Tanner},
and thermal conductivity $\kappa$ \cite{Gagnon97}.
Consistently a value of roughly {\em two}
is found for the ratio of the carrier density 
to the effective electron mass (or the square
of plasmon frequency) between the $b$-axis and the $a$-axis.
This implies, in regard to YBCO,
that the contribution from the CuO chain is crucial
and of the same order as the CuO$_2$ plane contribution. 

It is well known, following a Drude-like model, that
the normal-state DC electrical conductivity is given by
$\sigma_n=n e^2 \tau/m$ with $n$ the carrier density, $m$ the
effective mass of the carrier, and $\tau$ the average scattering time.
When the system undergoes a phase transition
into a superconducting $d_{x^2-y^2}$ state, 
a universal microwave conductivity $\sigma_s^0=n e^2 /\pi m\Delta_0$ 
($\Delta_0$ is the maximum gap on Fermi surface) 
is predicted at low temperatures \cite{Lee93}. This universality arises due to
the cancellation between the finite impurity-induced
quasiparticle density of states at zero energy and the scattering time.
Universal features have been confirmed in YBCO by Zhang {\em et al.} 
\cite{Zhang} for the microwave conductivity and by 
Taillefer {\em et al.} \cite{TLGBA97} for the thermal conductivity.
Clearly $\sigma_n$ and $\sigma_s^0$ share a common
dependence on $n/m$ so that one expects
a similar in-plane $a$-$b$ axis anisotropy for them.
Apart from the dependence on $n/m$,
an anisotropic $\sigma_n$ can also arise from 
an anisotropic scattering time $\tau$, while the universal 
$\sigma_s^0$, by the meaning of ``universal'', is
independent of the scattering rate.

More recent thermal conductivity
measurements of Chiao {\em et al.} \cite{Chiao98}
which confirm previous microwave conductivity measurements 
of Zhang {\em et al.} \cite{Zhang} on YBCO
have observed the predicted low-temperature universality \cite{Lee93}
in both $a$ and $b$ directions, but observed no significant anisotropy
between the two.
This is in contradiction with the large anisotropy observed
in many other quantities.
 
In this paper, we propose a scenario that resolves the discrepancy 
mentioned above.  In reality, YBCO has 
three conducting layers (two CuO$_2$ plane and one CuO chain layer) 
within a unit cell. There is strong evidence 
\cite{Atkinson96,Odonovan97-3,Odonovan97-2,Odonovan97-1} that
the gap in all three layers is mainly $d_{x^2-y^2}$-wave like with 
some small admixture of an $s$-wave component \cite{Sun,Kouznetsov}.
The sub-dominant component arises
because of the orthorhombic nature of YBCO.
The Fermi surfaces are quite different
in the different layers. For isolated layers, it is 
tetragonally two-dimensional for the two CuO$_2$ planes, while 
it is highly one-dimensional for the CuO chain layer. 

\begin{figure}[h]
\vspace{-0.2cm}
\postscript{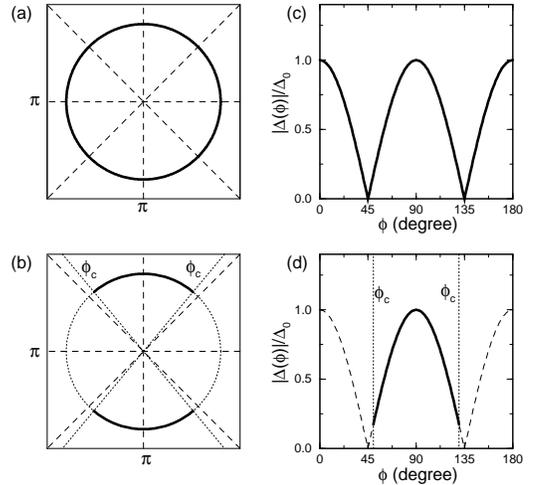}
\vspace{-0.7cm}
\caption{Fermi surface for CuO$_2$ plane (a) and CuO chain (b) layers
in YBCO. Frames (c) and (d) are the $d_{x^2-y^2}$-wave gaps seen on the
Fermi surfaces for the plane and chain layers.}
\label{fig1}
\end{figure}

A model that captures the physics in a simple way is the following.  
We assume all three layers in YBCO to
have a circular Fermi surface as is usually done
and assume a $d_{x^2-y^2}$-wave gap,
described by $\Delta_{\bf k}=\Delta_0 \cos(2\phi)$ as usual
with $\phi$ a polar angle in the 2D Brillouin zone.
But to model for the single chain layer, we introduce 
a cutoff angle $\phi_c$ 
(see Fig.~\ref{fig1}) which corresponds to an open circular Fermi surface 
and simulates the highly one-dimensional nature of the chain band. 
The Fermi surface for chain band is shown as in Fig.~\ref{fig1}(b)
with the CuO chain taken to be along the $k_y$ axis. 
A cutoff angle, $\phi_c>\pi/4$, corresponds to the angular position 
where a minimum gap starts to appear on the Fermi surface of the chain layer.
The measured properties will correspond to the
sum of the contributions from the two plane layers
(associated with a complete Fermi-surface integration) and 
the one chain layer (associated with an
incomplete Fermi-surface integration).
 
Measurements on YBCO can be divided into two categories.
The ones which are strongly dependent on seeing the 
gap nodes on Fermi surface ({\em e.g.}, $\sigma_s^0$)
and anothers which are not too dependent 
such as the value of the zero-temperature penetration depth $\lambda(0)$.  
In the former case, due to the cutoff angle,
the gap nodes may not cross anywhere on the entire Fermi surface for
the chain band, and anisotropy is consequently suppressed because 
of the reduction of universal conductivity in the chain layer.  

\begin{figure}[h]
\vspace{-0.5cm}
\postscript{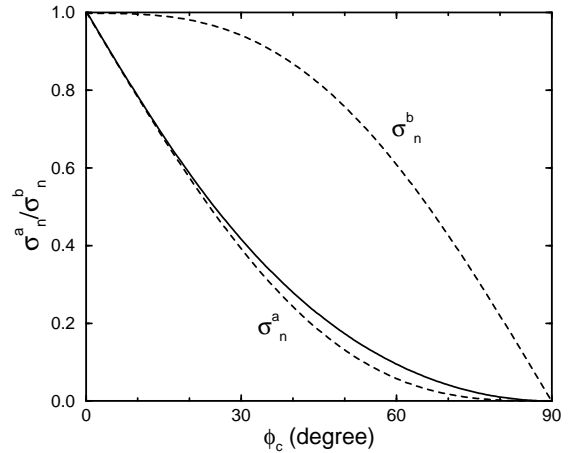}
\vspace{-0.2cm}
\caption{Solid line is the
ratio of $\sigma_n^a/\sigma_n^b$ as a function of cutoff
angle $\phi_c$ for the chain contribution. Dashed lines correspond
to $\sigma_n^a$ and $\sigma_n^b$ individually in arbitrary units.}
\label{fig2}
\end{figure}

{\em Normal-State Transport.}---
Assuming that the scattering rate is not dependent on the direction
of the momentum ${\bf k}$,
the ratio of normal {\em chain} conductivity along the $a$ and $b$
axis is then given by ($v$ is the Fermi velocity)

\begin{eqnarray}
{\sigma_n^a\over \sigma_n^b}
={\langle v_x^2 \rangle_c\over \langle v_y^2\rangle_c}=
{\langle \cos^2\phi\rangle_c\over \langle \sin^2\phi\rangle_c}, 
\label{eq:sigmaN.ratio}
\end{eqnarray}
where, $\langle A \rangle_c\equiv (2/\pi)\int_{\phi_c}^{\pi/2}d\phi~A$, 
denotes an incomplete Fermi-surface average
for the chain layer. In Fig.~\ref{fig2}, we plot the ratio,
as a function of the cutoff angle $\phi_c$, given in
Eq.~(\ref{eq:sigmaN.ratio}).
It is seen, as expected, that $\sigma_n^b > \sigma_n^a$ when
an incomplete Fermi-surface integration is carried out because 
$\langle v_y^2 \rangle_c$ has larger components than
$\langle v_x^2 \rangle_c$.
When $\phi_c$ is large corresponding to a flatter band,
$\sigma_n^b \gg \sigma_n^a$.
For example in the case $\phi_c=45^o$, $\sigma_n^b/\sigma_n^a\simeq 4$. 
This, in combination with the plane contribution,
conforms with the large observed anisotropy between the the $a$ and $b$-axis 
resistivities \cite{Friedmann90}.
In a strictly 1D model for the chains, $\sigma^a$ would be zero. Here
it is finite but becomes small when $\phi_c$
is large which approaches the chain case more closely
and the model is good enough for qualitative arguments.

Similar arguments can be applied to the anisotropy 
exhibited in the zero-temperature value of the 
{\em London penetration depth}.
Recent measurement of Basov {\em et al.} \cite{Basov}
showed for pure YBCO that $\lambda_a(0) \simeq 1600 {\rm \AA}$ and
$\lambda_b(0) \simeq 1000 {\rm \AA}$.
In the pure limit,
the square of the inverse penetration depth is given by
 
\begin{eqnarray}
{1\over \lambda_{\mu}^{2}}(T)={8\pi e^2\over c^2}{1\over \Omega}
\sum_{\bf k} v_{{\bf k},\mu}^2\left[
{\partial f(E_{\bf k})\over\partial E_{\bf k}}-
{\partial f(\epsilon_{\bf k})\over\partial \epsilon_{\bf k}} \right]
\label{eq:pene.depth}
\end{eqnarray}
and is associated with the superfluid density.
Here $e$ is electron charge, $c$ the velocity of light,
$v_{{\bf k},\mu}$ is $\mu$ component of the Fermi velocity,
and $f$ the Fermi-Dirac distribution function.
The electron energy in the normal state is $\epsilon_{\bf k}$ and
$E_{\bf k}$ is the quasiparticle energy in the superconducting state with
$E_{\bf k}=\sqrt{\epsilon_{\bf k}^2+\Delta_{\bf k}^2}$.
The zero-temperature value of 
$\lambda_\mu$ is determined solely by the
second term in (\ref{eq:pene.depth})
and depends only on normal-state parameters.
By considering an incomplete Fermi-surface integration for
the chain layer, one obtains
 
\begin{eqnarray}
{\lambda_b^2(0)\over \lambda_a^2(0)}
={\langle v_x^2 \rangle_c\over \langle v_y^2\rangle_c}=
{\langle \cos^2\phi\rangle_c\over \langle \sin^2\phi\rangle_c}.
\label{eq:lambda.ratio}
\end{eqnarray}
Thus, similar to Eq.~(\ref{eq:sigmaN.ratio}) for the resistivity ratio,
the introduction of a cutoff angle will naturally
lead to the right ratio for the observed zero-temperature anisotropy of
the penetration depths. (One must add in a plane contribution
when comparison with experiment is made.) 

{\em Superconducting-State Transport.}--- Using the Kubo formalism,
the zero-temperature microwave conductivity along the 
$\mu$-direction for the chain layer is given by 

\begin{eqnarray}
\sigma_{\mu}(0)=e^2\sum_{\bf k}^c\left[
{v_\mu^2 \gamma^2\over (\gamma^2+\epsilon_{\bf k}^2+
\tilde{\Delta}_{\bf k}^2)^2}
\right],
\label{eq:sigma}
\end{eqnarray}
where $c$ denotes an incomplete Fermi surface average,
$\gamma=i\Sigma_0(\omega=0)$ is the impurity-induced scattering rate,
and $\tilde{\Delta}_{\bf k}=\Delta_{\bf k}+\Sigma_1(\omega=0)$
is the impurity renormalized gap at zero frequency with 
$\Sigma_i$ the self-energies (see later).
It is known that, for a gap with $d_{x^2-y^2}$-symmetry and a
complete Fermi surface average, there will be no
renormalization effect on the gap ($\Sigma_1=0$), 
but this is no longer the case for a chain. After carrying out the 
energy integration, Eq.~(\ref{eq:sigma}) is reduced to

\begin{eqnarray}
\sigma_{\mu}(0)=N(0)e^2\left\langle
{\gamma_\mu^2(\phi)
\gamma^2\over [\gamma^2+\tilde{\Delta}^2(\phi)]^{3\over 2}}\right\rangle_c,
\label{eq:sigma0}
\end{eqnarray}
where $N(0)=m/(2\pi\hbar^2)$ is the electronic
density of states at the Fermi surface and $m$ is the electron
mass for the band. For the $a$-axis, $\gamma_a(\phi)=\cos\phi$ and 
for the $b$-axis, $\gamma_b(\phi)=\sin\phi$.

\begin{figure}[h]
\vspace{-0.5cm}
\postscript{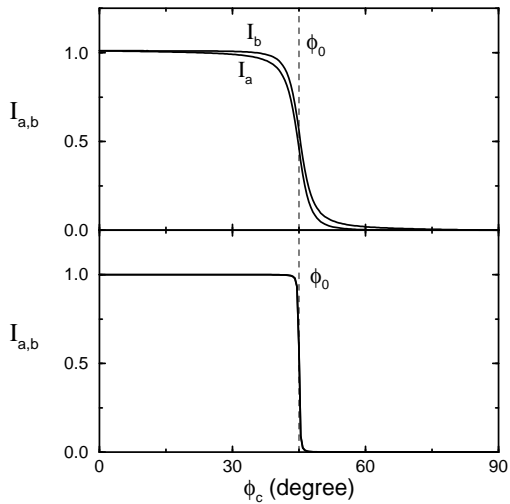}
\vspace{-0.2cm}
\caption{Values of $I_a$ and $I_b$ defined in
Eq.~({\protect\ref{eq:Iab}}) in terms of
the cutoff angle $\phi_c$. The angle $\phi_0$ corresponds to the node.
Top frame is for
$\gamma/\Delta_0=0.1$ and bottom frame is for $\gamma/\Delta_0=0.01$.}
\label{fig3}
\end{figure}

To show the physics in a transparent way, we first ignore
the gap renormalization, 
$\tilde{\Delta}(\phi)\rightarrow \Delta(\phi)=\Delta_0 \cos(2\phi)$,
and manipulate  Eq.~(\ref{eq:sigma0}) to obtain
 $\sigma_{a,b}(0)\equiv (ne^2/\pi m\Delta_0) I_{a,b}(\phi_c,\gamma)$, where
$n=k_F^2/2\pi$ is the layer carrier density and 

\begin{eqnarray}
I_{a,b}(\phi_c,\gamma)\equiv 2\int_{\phi_c}^{\pi/2}d\phi
{(\cos^2\phi, \sin^2\phi)\bar{\gamma}^2 \over 
[\bar{\gamma}^2+\cos^2 2\phi]^{3\over 2}}.
\label{eq:Iab}
\end{eqnarray}
Here, $\bar{\gamma}\equiv \gamma/\Delta_0$ is the normalized scattering rate.
The effect of impurities on the renormalization of the gap
and on the effective scattering rate will be discussed later.
In Fig.~\ref{fig3}, we plot $I_{a,b}$ 
for two different choices of $\bar{\gamma}$,
namely $\bar{\gamma}=0.1$ and $0.01$.
For small $\gamma$, the function $I_a$ is indistinguishable from $I_b$
at all values of $\phi_c$. Only when $\gamma$ is large,
does $I_b$ starts to deviate from $I_a$ and,
due to the fact that the chain Fermi surface mainly develops
in the $\phi=\pm \pi/2$ region and has no contribution from the
$\phi=0,\pi$ region, $\sigma_{b}(0)$ is slightly 
larger than $\sigma_{a}(0)$.
More importantly, $I_{a,b}$ almost vanish as long as $\phi_c>\phi_0$
(nodal angle) in the small $\gamma$ case.
(In our present case for simplicity, we have $\phi_0=\pi/4$.) 

The small saturated low-temperature
microwave conductivity in a $d_{x^2-y^2}$-wave superconductor 
is recovered from Fig.~\ref{fig3} in the case of $\phi_c \rightarrow 0$.
As is clearly seen in Fig.~\ref{fig3}, 
both $I_{a}$ and $I_b$ approach one and the conductivity is saturated
to the value $\sigma_{a}(0)=\sigma_{b}(0)\rightarrow 
\sigma_s^0=ne^2/(\pi m\Delta_0)$, independent
of $\gamma$, {\em i.e.}, of the impurity concentration.
It is worth emphasizing that both $\sigma_{a}(0)$ and $\sigma_{b}(0)$
are close to the saturated value $\sigma_s^0$ as long as $\phi_c< \phi_0$. 
The latter case means that the gap nodes cross pieces of the Fermi surface.  
In the opposite case $\phi_c>\phi_0$, 
the residual conductivity $\sigma_{a}(0)$ and $\sigma_{b}(0)$ 
(or $I_a$ and $I_b$) are {\em both} strongly suppressed.
In this case, no gap nodes cross the one-dimensional chain Fermi surface. 
This is an effect due solely 
to the geometry of the chain Fermi surface and arises 
even though the gap used in our model calculation 
has pure $d_{x^2-y^2}$ symmetry.
Moreover, when $\phi_c> \phi_0$ and a minimum gap 
$\Delta_{\rm min}\equiv \Delta_0 |\cos(2\phi_c)|$ is seen on the
chain Fermi surface,
one can show that $I_{a}< \pi \langle \cos^2\phi\rangle_c
\bar{\gamma}^2/(\bar{\gamma}^2+\bar{\Delta}_{\rm min})^{3/2}$
and  $I_{b}< \pi \langle \sin^2\phi\rangle_c
\bar{\gamma}^2/(\bar{\gamma}^2+\bar{\Delta}_{\rm min})^{3/2}$.
Here $\bar{\Delta}_{\rm min}\equiv \Delta_{\rm min}/\Delta_0$.
This means that both $I_a$ and $I_b$ are negligibly 
small when $\gamma\ll \Delta_{\rm min}$.

Thus, for the case of YBCO under consideration here,
there is a possibility that
the gap nodes do not cross the entire Fermi surface
for the highly one-dimensional chain band ($\phi_c > \phi_0$) 
and, as a consequence, no saturation for $\sigma_{a,b}(0)$ occurs
in the chain layer.  In addition, for the same reason that
the residual conductivity is strongly suppressed
in chain layer (for both $a$ and $b$-axes), there is
no significant anisotropy in the residual microwave conductivity.
 
In order to understand what the temperature regime is, in which
the universal value is valid for YBCO, one recalls that
the low-temperature microwave conductivity,
$\sigma(T)\simeq\sigma_s^0(1+T^2/\gamma^2)$, as given by
Hirschfeld {\em et al.} \cite{HPS} for a 
$d_{x^2-y^2}$-wave superconductor with a complete circular Fermi-surface.
Therefore, for the CuO$_2$ planes only, one needs to have $T\ll \gamma$
to achieve universality. For the chain layer 
in which the gap nodes do not cross the chain-band Fermi surface,
a minimum gap $\Delta_{\rm min}$
is seen and a result similar to the $s$-wave
case holds for the low temperature residual conductivity, namely
$\sigma(T)\simeq\sigma_n (\Delta_{\rm min}/T) 
e^{-\Delta_{\rm min}/T}$ is predicted (see, for example, Ref.~\cite{HPS}).
The chain residual conductivity is thus
strongly suppressed until
the temperature $T\agt \Delta_{\rm min}$.
The above picture corresponds exactly to the strong suppression
of $\sigma_a(0)$ and $\sigma_b(0)$ for chain layer discussed before.
Hence, the chain layer will have little effect on the
temperature limit where universality holds
if $\Delta_{\rm min}$ associated with the chain is larger than 
$\gamma$ associated with the plane.

\begin{figure}[h]
\vspace{-0.5cm}
\postscript{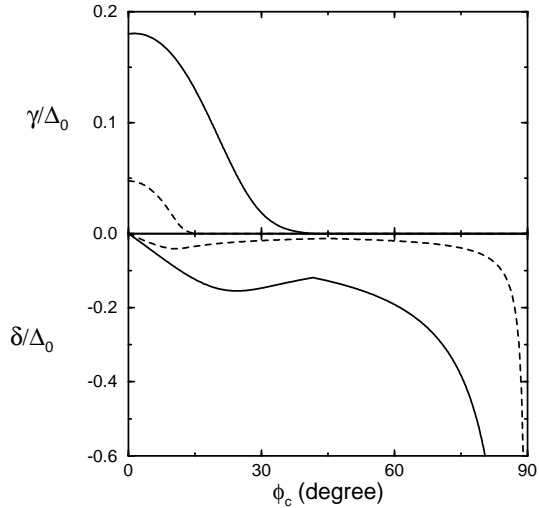}
\vspace{-0.2cm}
\caption{The impurity-induced
effective scattering rate $\gamma$ and gap renormalization $\delta$
scaled to $\Delta_0$ (gap maximum in the absence of impurity). 
Solid line is for $n_i/4N(0)\Delta_0=0.1$ and
dashed line is for $n_i/4N(0)\Delta_0=0.01$
with $n_i$ the impurity concentration.}
\label{fig4}
\end{figure}

Now we discuss the effect of impurity
scattering on the zero-frequency effective scattering rate 
[$\gamma=i\Sigma_0(\omega=0)$] and gap renormalization
[$\delta\Delta_{\bf k}=\tilde{\Delta}_{\bf k}-\Delta_{\bf k}=
\Sigma_1(\omega=0)$].
In the usual $t$-matrix approximation, the self-energies 
$\Sigma_0(\omega)=n_i G_0/(c^2- G_0^2+ G_1^2)$
and $\Sigma_1(\omega)=-n_i G_1/(c^2- G_0^2+ G_1^2)$
with $n_i$ the impurity concentration, $c^2\equiv 1/v_i^2$
($v_i$ is the impurity potential), and the integrated Green's function 
$G_i(\omega)\equiv \sum_{\bf k}G_i({\bf k},\omega)$.
We note in general that the impurity renormalization effect should also be
included in the band structure $\delta\epsilon_{\bf k}=\Sigma_3$
(in a general sense, $\phi_c$ will also be renormalized in our model).
However, this is safely neglected.  

In the unitary resonant scattering limit ($c=0$), we obtain
the coupled equations ($\delta\Delta_{\bf k} \equiv \delta$)

\begin{eqnarray}
\gamma&=&{n_i\over 2\pi N(0)}{I_0(\phi_c)\over I_0^2(\phi_c)+I_1^2(\phi_c)}
\nonumber\\
\delta&=&{n_i\over 2\pi N(0)}{I_1(\phi_c)\over I_0^2(\phi_c)+I_1^2(\phi_c)},
\label{eq:gamma.delta.1}
\end{eqnarray}
where

\begin{eqnarray}
I_0(\phi_c)&=&\left\langle{\bar{\gamma}\over \sqrt{\bar{\gamma}^2
+(\cos 2\phi+\bar{\delta})^2}}\right\rangle_c
\nonumber\\
I_1(\phi_c)&=&\left\langle{\cos 2\phi+\bar{\delta}\over \sqrt{\bar{\gamma}^2
+(\cos 2\phi+\bar{\delta})^2}}\right\rangle_c
\label{eq:gamma.delta.2}
\end{eqnarray}
with $\bar{\delta}\equiv \delta/\Delta_0$.
In Fig.~\ref{fig4}, we show our self-consistent solutions for $\gamma$ and 
$\delta$ as a function of $\phi_c$ given by Eqs.~(\ref{eq:gamma.delta.1})
and (\ref{eq:gamma.delta.2}). 
We have used two different impurity concentrations, namely
$n_i/4N(0)\Delta_0=0.1$ and $0.01$.
Two features are noted in Fig.~\ref{fig4}: (i) $\gamma$ is strongly suppressed
in the large $\phi_c$ case and (ii) the gap renormalization $\delta$
is always negative. The strong suppression of $\gamma$ in 
the larger $\phi_c$ case is a natural consequence 
of the evolution of a $d$-wave superconductor towards an
$s$-wave superconductor ({\em i.e.}, a minimum gap $\Delta_{\rm min}$
develops in the chain). 
On the other hand, the result that the gap renormalization $\delta$
is negative is because the highly 1D Fermi surface
samples mainly the negative portion of the $d$-wave gap.

Analytical results are available in three different regimes.
In the $\phi_c\rightarrow 0$ limit, one recovers the results for
the complete Fermi surface case (appropriate for CuO$_2$ plane) that is
$\gamma\sim (\Delta_0/\tau)^{1/2}$ [$\tau^{-1}\sim n_i/N(0)$ is
the normal-state scattering rate] and $\delta=0$
({\em i.e.}, no gap renormalization). When $\phi_c$ is small but finite
(in the case that Fermi surface still sees nodes but the
band structure has evolved from the tetragonal to the orthorhombic limit),
we find $I_0\simeq -\bar{\gamma}\ln \bar{\gamma}$ and
$I_1\simeq  -1/\phi_c$. Consequently,
$\gamma\sim \Delta_0\exp(-\tau\Delta_0\phi_c^2)$ and
$\delta\sim -1/(\tau\phi_c)$.
Thus, $\gamma$ is decaying exponentially as $\phi_c$ increases
and $|\delta|$ is proportional to $1/\phi_c$. 
When $\phi_c>\phi_0$ such that a minimum gap $\Delta_{\rm min}$
develops on the Fermi surface,
we find $\gamma$ is small (similar to the $s$-wave case),
while $\delta\sim -1/[\tau(\pi/2-\phi_c)]$.
Therefore, $|\delta|$ increases as $\phi_c$ increases and
diverges when $\phi_c$ approaches $\pi/2$. Of course, the latter
case is an unphysical regime. 
The three different regimes for $\delta$ are seen most clearly in Fig.~\ref{fig4}.

Since $\delta<0$,
the gap nodes will be effectively shifted downward (or equivalently
$\phi_0<\pi/4$).  As a result, this effect increases the
chance that the gap nodes will not cross the entire chain Fermi surface.
In addition to the effect of impurity, 
the appearance of a {\em negative} $s$-wave admixture in the 
$d_{x^2-y^2}$-wave gap \cite{Sun,Kouznetsov}
can naturally distort the gap nodes to be less than
45 degree and thus leads to a similar consequence.
A finite minimum gap on the chain Fermi surface is therefore
seen to be a robust feature of the one dimensional nature of the bands 
and of a $d$-wave gap.
Recently, a detailed survey of the effect of Ni impurity (substitution
for Cu) on YBCO has been carried out by Homes {\em et al.}
\cite{Homes98} using optical measurement. They concluded that the 
deposition of Ni is mainly situated in the chain layer.
Hence the change with Ni impurity concentration can be used
to test our model.

W.C.W. acknowledges the support from 
National Science Council (NSC) of Taiwan under Grant No.
NSC 87-2112-M-003-014 and the hospitality of
National Center of Theoretical Science (CTS) of Taiwan. 
J.P.C. acknowledges the support from Natural Sciences 
and Engineering Research Council (NSERC) of Canada
and of the Canadian Institute for Advanced Research (CIAR).

\end{document}